\documentclass[twocolumn,showpacs,preprintnumbers,amsmath,amssymb]{revtex4}
\usepackage[dvips]{graphicx}
\usepackage{dcolumn}% Align table columns on decimal point
\usepackage{bm}% bold math

\begin{document}

\preprint{}

\title{On the Dual Structure of Thermodynamics}

\author{A. Porporato}
\email{amilcare@duke.edu}
 \affiliation{Department of Civil and
Environmental Engineering, Duke University, Durham NC, USA}

\date{\today}

\begin{abstract}
Based on the properties of exponential distribution families we analyze the Fisher information of the Gibbs canonical ensemble to construct a new state function for simple systems with no mechanical work. Such a function possesses nice symmetry properties with respect to Legendre transform and provides a connection with previous alternative formulations of thermodynamics, most notably the work by Biot, Serrin and Frieden and collaborators. Logical extensions to systems with mechanical work may similarly consider generalized Gibbs ensembles.\end{abstract}

%\pacs{02.50.-r, 05.40.-a}% PACS, the Physics and Astronomy

\maketitle

\section{Introduction}

The theory of classical thermodynamics elegantly links energy and matter states in macroscopic systems in thermodynamic equilibrium  \cite{Callen,Bejan}. It also allows us to rigorously describe the overall effect of transformations between different states of equilibrium, although it does not provide information about their actual dynamics. Historically, thermodynamics developed from the intuitive notions of heat and temperature, while subsequently it became apparent the existence of more abstract, fundamental quantities such as internal energy, entropy and their corresponding Legendre transforms, which are now considered primitive concepts in axiomatic and postulational theories of thermodynamics. However, considerable effort has also been devoted to rigorously reformulate thermodynamics starting from temperature and heat as primitive quantities. We refer in particular to the work by Serrin and Silhavy \cite{Ser1,Ser2,Sil1,Sil2}, who derived internal energy and entropy starting from an analytical formulation of Clausius integral for cyclic processes based on temperature and a newly defined heat accumulation function. Related theories were also proposed in \cite{Green,Lieb}.

The inability of classical thermodynamics to describe nonequilibrium dynamics also spurred the search for alternative field theories of thermodynamics. In particular, Onsager's theory \cite{Callen}, based on local equilibrium and a variational principle for a dissipation function (proportional to entropy production), helped explain the coupling among processes for small deviations from thermodynamics equilibrium. However, as it has been repeatedly pointed out \cite{Prig,Gyar,Cafa,Callen,GuoPRSA}, a variational formulation based on entropy production extremization does not lead to the observed linear phenomenological laws commonly used in applications (e.g., the Fourier law of heat conduction has constant conductivity, while entropy production maximization leads to a heat conductance which is inversely proportional to the square of temperature). Several attempts were thus pursued to find alternative variational formulations. The work of Biot \cite{Biot1,Biot2} in particular showed that a variational formulation based on a quadratic function of temperature, obtained from a suitable linearization of thermodynamic availability for constant heat capacity, provides Fourier's heat conductance law. The theoretical bases of Biot's function were not completely clarified, however, and the extension to compressible media remained elusive. More recently, Guo and coworkers \cite{Guo07,GuoPRSA} resumed Biot's function and used a similar quantity, called entransy and justified on the bases of a formal analogy between heat and electrical currents, to optimize engineering design in heat transfer.

Independently, a series of remarkable contributions by Frieden and collaborators \cite{Fri89,Fri90,Friso95,Friebook,Frietal99} showed that variational formulations based on Fisher information may serve as a powerful means to obtain several physical laws, at the same time providing a reason for the ubiquitous presence of squares of gradients in such laws. Moreover, the additivity property of the Fisher information and its capability to measure 'disorder' \cite{Friebook} are particularly appealing features from a thermodynamic point of view. More importantly, Frieden et al. \cite{Frietal99,Friebook} showed that Fisher information is endowed with a Legendre-transform structure which mirrors that of classical thermodynamics. This contribution however was largely formal and the link with the usual thermodynamic variables remained somewhat unexplored. In an interesting step in this direction \cite{Penn}, much like Biot's function Fisher information too was found to be a quadratic function of temperature for the case of constant heat capacity, with a proportionality coefficient containing an unspecified reference temperature.

In what follows we discuss how these previous lines of research may be connected and show how a dual structure of thermodynamics emerges from the the formalism of exponential families (to which Gibbs equilibrium distributions belong) and their link to Legendre and cumulant transforms \cite{Band,Cox,Dav}. With specific reference to simple system with constant volume and mass, we show how the correspondence between thermodynamic potentials, in either the entropy or energy representation \cite{Callen}, is mirrored by an equivalent, dual representation based on Fisher information. As pointed out by the work of Biot \cite{Biot2} and Frieden and coworkers \cite{Friebook}, this dual structure may be useful in variational formulations of nonequilibrium field theories of thermodynamics and continuum media.

\section{Canonical Ensemble and Fisher Information}

Consider a simple system with fixed volume $V$ and particle number $N$ in thermal equilibrium with a reservoir at temperature $T=1/\beta$ (to simplify notation, the Boltzmann constant is equal to one so that temperature is in natural units). For our purposes here, it is convenient to write the Gibbs canonical distribution for a single random variable representing the internal energy, $\hat{U}$,
\begin{equation}\label{eq:Gibbsdistr}
p(\hat{U};\beta)=\frac{1}{Z(\beta)} e^{-\beta \hat{U}-h(\hat{U})}=e^{\Phi(\beta)-\beta \hat{U}-h(\hat{U})},
\end{equation}
where $Z(\beta)$ is the partition function, $h(\hat{U})$ is a function linking the microstates to the internal energy, and the Massieu function is $\Phi(\beta)=\ln Z(\beta)=-F/T$, with $F$ the Helmholtz free energy.  It is also useful to write the distribution in terms of the fluctuations of internal energy around its mean value, $\hat{U'}=\hat{U}-U$,
\begin{equation}\label{eq:Gibbsdistr1}
p(\hat{U}';\beta)=e^{S(\beta)- \beta \hat{U}'-h'(\hat{U}')},
\end{equation}
where $S(\beta)=\Phi+\beta U$ is the equilibrium entropy expressed as a function of $\beta$.

The probability density function (\ref{eq:Gibbsdistr}) belongs to the exponential family \cite{Band,Cox,Dav}, for which several interesting properties are known. In particular, $-\Phi$ and $-S$ are the cumulant transforms \cite{Cox} of the respective distributions for $\hat{U}$ and $\hat{U}'$ \cite{Note3}, so that
\begin{equation}
-\frac{d\Phi}{d\beta}=\langle \hat{U}\rangle = U.
\end{equation}
More importantly, their second derivatives are also the Fisher information of the distribution with respect to $\beta$ as well as the variance of $\hat{U}$,
\begin{equation}\label{eq:varU1}
\mathcal{F}_U(\beta)=-\frac{d^2 \Phi}{d \beta^2}=-\frac{d^2 S}{d\beta^2}= \langle (\hat{U}-U)^2\rangle,
\end{equation}
which is
\begin{equation}\label{eq:varU2}
\mathcal{F}_U(\beta)=-\frac{dU}{d\beta}=\frac{N c_v(\beta)}{\beta^2}=T^2 N c_v(T),
\end{equation}
where $c_v(\beta)$ is the specific molar heat capacity. Such a result is in agreement with the well established theory of fluctuations (see \cite{Callen2} and in particular Eq. (19.6) in \cite{Callen1}). Recall, in general, that the Fisher information is defined as the variance of the score, \begin{eqnarray}\label{eq:fisherdef1}
\mathcal{F}_X(\theta) & = &-\int_x \left(\frac{\partial \ln p(x|\theta)}{\partial \theta}\right)^2 p(x|\theta)dx\\
 &=&-\int_x\frac{\partial^2 \ln p(x|\theta)}{\partial \theta^2} p(x|\theta)dx,\nonumber
\end{eqnarray}
where the second form holds under suitable regularity conditions met here.

Eq. (\ref{eq:varU2}) shows that the Fisher information corresponds, for constant heat capacity, to the function used by Biot \cite{Biot1,Biot2} and Guo et al. \cite{Guo07,GuoPRSA} in their variational formulation of heat conduction, as well as to the quantity obtained by \cite{Penn} apart from the arbitrary temperature. The theory of exponential families allows also to make immediate contact with the stability theory of equilibrium conditions (see e.g. \cite{Callen}, Eq. (8.7)). Considering in fact that $S$ is the log-likelihood function of $\hat{U}$ and the Legendre transform of $\Phi$, it follows that (\cite{Cox}, theorem 6.1, Eq. (6.42))
\begin{equation}\label{eq:stab1}
\frac{dS}{dU}=\beta,
\end{equation}
and
\begin{equation}\label{eq:stab2}
\frac{d^2S}{dU^2}=\frac{d\beta}{dU}=-\frac{\beta^2}{N c_v}=-\frac{1}{ N c_v T^2}=-\mathcal{F}_U^{-1}\le 0.
\end{equation}
Eq. (\ref{eq:stab2}) has been used as a measure of the curvature in a Riemanninan representation of the thermodynamic space \cite{Rup79,Brod}, and is linked to the availability (i.e. free energy differences with respect to equilibrium) of small perturbations around equilibrium \cite{Salo} (a version of the Gouy-Stodola theorem \cite{Bejan}).

For the system at constant $V$ and $N$, $\mathcal{F}_U$ is an extensive quantity, which may be related to the ability to measure the inverse temperature $\beta$ of the reservoir by observing the state of internal energy of the system. Such an ability increases with the size of system $N$, because the fluctuations get smaller with $N$, but also increases with $T^2$ proportionally to the heat capacity $c_v(T)$ at that temperature, because larger heat capacities buffer temperature fluctuations which would otherwise impair the estimate. This idea of using the system as temperature measurement device can be traced back to \cite{Mand,Tisza}.

\section{Dual Structure of Thermodynamics}

The first equality in (\ref{eq:varU2}),
\begin{equation}\label{eq:varU3}
\mathcal{F}_U(\beta)=-\frac{dU}{d\beta},
\end{equation}
underlies another Legendre transformation, implying (\cite{Callen} p. 142) that $U(\beta)$ is related by Legendre transform to
\begin{equation}\label{eq:inv1}
\mathcal{F}_S=\beta \mathcal{F}_U+U.
\end{equation}
Although the previous equation is in units of energy, it is formally analogous to $S=\beta U+\Phi$ in the familiar entropy representation of thermodynamics. When expressed as a function the internal energy, $\mathcal{F}_U(U)$ may be used as fundamental equation in alternative to $S(U)$, for systems with no reversible (volume) work and mass exchanges.
It is also natural to consider a dual version of the energy representation by dividing the previous equation by $\beta$ and re-expressing it in terms of $T$,
\begin{equation}\label{eq:inv2}
\mathcal{F}_U=T\mathcal{F}_S+\mathcal{F}_F,
\end{equation}
where $\mathcal{F}_F=UT$. This equation corresponds to the familiar $U=TS+F$ in the energy representation.

Considering now the differential of (\ref{eq:inv1})
\begin{equation}
d\mathcal{F}_S=d \beta \mathcal{F}_U+\beta d\mathcal{F}_U+dU,
\end{equation}
and that, using (\ref{eq:varU3}) and considering that in the absence of reversible work changers in internal energy are equal to reversible heat flow, $dQ^{rev}$,
\begin{equation}
dU=-d\beta \mathcal{F}_U=dQ^{rev}=TdS,
\end{equation}
so that
\begin{equation}
d\mathcal{F}_U=T d\mathcal{F}_S=T^2 dS.
\end{equation}
This relation has two interesting consequences. Firstly, since $d \mathcal{F}_S=TdS=dU$, then at constant $V$ and $N$, $\mathcal{F}_S=U+ const$, thus pointing to a Legendre-transform invariance between the couples $\mathcal{F}_S$ and $U$ in (\ref{eq:inv1}) and $\mathcal{F}_U$ and $\mathcal{F}_F$ in (\ref{eq:inv2}). Such an invariance is immediately evident if one considers the case of constant heat capacity in which $\mathcal{F}_U$ expressed in terms of $T$ is a quadratic function of temperature, while its Legendre transform is still a parabola but now in terms of $U$.

Secondly, because of (\ref{eq:varU2}),
\begin{equation}\label{eq:av1}
\frac{d\mathcal{F}_U}{dS}=T^2=-\frac{\mathcal{F}_U}{N c_v},
\end{equation}
so that
\begin{equation}\label{eq:av2}
\Delta \ln \mathcal{F}_U=-\int \frac{dS}{Nc_v}
\end{equation}
and, if the heat capacity is constant ($c_v=c$),
\begin{equation}\label{eq:av3}
\Delta \ln \mathcal{F}_U=-\frac{\Delta S}{Nc}.
\end{equation}
The case of constant heat capacity corresponds to energy equipartition(\cite{Callen}, p. 291 and 376), where the internal energy has a quadratic dependence on the (micro)states of the $\nu=2c$ degrees of freedom such that the resulting canonical distribution (\ref{eq:Gibbsdistr}) is a chi-quare (gamma) distribution with $N\nu$ degrees of freedom and scale parameter $\beta$, mean $c/\beta$ and variance $c/\beta^2$, while $U=NcT$, $S=Ns_0+Nc\ln T$ and $\mathcal{F}_U=NcT^2$. Considering the classical process of heat flow between two bodies of equal mass and heat capacity, initially at equilibrium at different temperatures (e.g. \cite{Callen} p. 101), it is easy to show that while heat conduction always increases the entropy, it always decreases $\mathcal{F}_U$ when approaching equilibrium at $T$.

We conclude by discussing the link of the present analysis with Serrin's theory \cite{Ser1,Ser2,Lampinen}, which provides an alternative formulation of thermodynamics based on hotness (i.e., temperature) and heat flow, without assuming internal energy and entropy as primitive concepts (similar concepts were developed by Silhavy \cite{Sil1,Sil2}; see also \cite{Lieb}). The theory is formulated using an accumulation function, $A(T)$, defined for cyclic processes as the total heat added in a process at temperature lower or equal than $T$,
\begin{equation}
\oint \frac{dQ}{T}=\int_0^\infty \frac{A(T)}{T^2}dT\leq0.
\end{equation}
Based on this definition, $A(T)$ is zero in adiabatic processes as well as whenever input and output of heat cancel at every temperature, while it is a step function for isothermal process. Considering an infinitesimal reversible transformation, it follows that
\begin{equation}\label{eq:serr2}
dS=\frac{dQ^{rev}}{T}=\frac{A(T)}{T^2}dT=\frac{d\mathcal{F}_U}{T^2},
\end{equation}
where $\mathcal{F}_U=\frac{dA}{dT}$. Thus, since $dA=\mathcal{F}_U dT$, $\mathcal{F}(T)$ can be seen as the density of heat accumulation in an infinitesimal process taking place between $T$ and $T+dT$.

\section{Conclusions}

The previous considerations where limited to the case of heat processes at constant volume and mass in simple systems. A logical way to extend the analysis to processes including mechanical work or mass flow would consider generalized Gibbs ensembles along with their Fisher information matrix. Such an extension and its consequences for field theories of thermodynamics will be explored in future contributions.

\section{Acknowledgements}
We thank A. Bejan for valuable discussions. This work was supported in part by NSF grant CBET-1033467.


\begin{thebibliography}{9}
\bibitem{Callen} H.B. Callen, {\it Thermodynamics and an Introduction to Thermostatistics}, J. Wiley \& Sons, New York (1965).
\bibitem{Bejan} Bejan A., {\it Advanced Engineering Thermodynamics} (3rd Ed.), Wiley, Hoboken NJ (2006).
\bibitem{Ser1} J. Serrin, Arch. Rat. Mech. Anal. {\bf 70}(4), 355 (1979).
\bibitem{Ser2} J. Serrin, An outline of thermodynamical structure, in
{\it New Perspective in Thermodynamics}, 3-31,  Springer, Heidelberg (1986).
\bibitem{Sil1} M. Silhavy,  Arch. Rat. Mech. Anal. {\bf 81}(4), 221 (1983).
\bibitem{Sil2} M. Silhavy, Foundations of continuum thermodynamics, in {\it New Perspective in Thermodynamics}, 33-47,  Springer, Heidelberg (1986).
\bibitem{Lieb} E.H. Lieb, J. Yngvason, Phys. Today {\bf 53}(4), 32-37 (2000).
\bibitem{Green} A.E. Green and P.M. Naghdi, Proc. R. Soc. Lond. A {\bf 432}, 171 (1991).
\bibitem{Prig} I. Prigogine, {\it Introduction to Thermodynamics of Irreversible Processes}, 3rd Ed. Interscience, New York (1967).
\bibitem{Gyar} I. Gyarmati, {\it Non-equilibrium Thermodynamics: Field Theoreis and Variational Principles}, Springer, New York (1970).
\bibitem{Cafa} V. Bertola and E. Cafaro, Int. J. Heat and Mass Transf. {\bf 51}, 1907 (2007).
\bibitem{GuoPRSA} Q. Chen, H. Zhu, N. Pan, and Z.-Y. Guo, Proc. Royal Soc. Lond. A {\bf 467}, 1012 (2011).
\bibitem{Biot1} M.A. Biot, J. Appl. Phys. {\bf 27}(3), 240 (1956).
\bibitem{Biot2} M.A. Biot, {\it Variational Principles in Heat Transfer}, Claredon Press, Oxford (1970).
\bibitem{Guo07} Z.-Y. Guo, H. Zhu, and X. Liang, Int. J. Heat and Mass Transf. {\bf 50}, 2545 (2007).
\bibitem{Fri89} B.R. Frieden, Am. J. Phys. {\bf 57}, 1004 (1989).
\bibitem{Fri90} B.R. Frieden, Phys. Rev. A {\bf 41}(8), 4265 (1990).
\bibitem{Friso95} B.R. Frieden and B.H. Soffer, Phys. Rev. E {\bf 52}, 2274 (1995).
\bibitem{Friebook} B.R. Friden, {\it Science from Fisher Information: a Unification}, Cambridge University Press, Cambridge (2004).
\bibitem{Frietal99} B.R. Frieden, A. Plastino, A.R. Plastino, B.H. Soffer, Phys. Rev. E {\bf 60}, 48 (1999).
\bibitem{Penn} F. Pennini, A. Plastino, Phys. Rev. E, {\bf 71},047102 (2005).
\bibitem{Band} O.E. Barndorff-Nielsen, {\it Information and Exponential Families in Statistical Theory}, J. Wiley \& Sons, Chichester (1978).
\bibitem{Cox} O.E. Barndorff-Nielsen, D.R. Cox, {Asymptotic Techniques for Use in Statistics}, Chapman and Hall London (1989).
\bibitem{Dav} A. Davison {\it Statistical Models}, Cambridge Univ. Press, Cambridge (2000).
\bibitem{Note3} The cumulant generating function of (\ref{eq:Gibbsdistr}) is actually $\kappa(a)=-\Phi(a-\beta)+\Phi(\beta)$; see e.g. \cite{Cox}.
\bibitem{Callen1} R.F. Green and H.B. Callen, Phys. Rev. {\bf 83}, 1231 (1951).
\bibitem{Callen2} H.B. Callen, Am. J. Phys. {\bf 33}(11), 919 (1965).
\bibitem{Salo} P. Salamon, J. Nulton, and E. Ihrig, J. Chem. Phys. {\bf 80}(1), 436 (1984).
\bibitem{Rup79} G. Ruppeiner, Phys. Rev A {\bf 79}(4), 1608 (1979).
\bibitem{Brod} D. Brody and N. Rivier, Phys. Rev. E {\bf 51}(2), 1006 (1995).
\bibitem{Mand} B. Mandelbrot, Ann. Math. Statist. {\bf 33} 1021 (1962).
\bibitem{Tisza} L. Tisza and P.M. Quay, Ann. Phys. {\bf 25}, 48 (1963).
\bibitem{Lampinen} M.J. Lampinen and J. Vuorisalo, J. Appl. Phys. {\bf 69}(2), 597 (1991).

\end{thebibliography}
\end{document}